# Development of a high-resolution and high efficiency Single Photon detector for studying cardiovascular diseases in mice

F. Garibaldi, E. Cisbani, F. Cusanno, F. Giuliani, M. Lucentini, G. Marano, F. Santanvenere
Istituto Superiore di Sanità, Rome, Italy

M. L. Magliozzi
e-Geos, Rome, Italy

P. Musico
Istituto Nazionale Fisica Nucleare, Sezione di Genova, Genova, Italy

G. De Vincentis,
Sapienza University, Rome, Italy

S. Majewski,
UC DAVIS University, CA, USA

Y. Wang and B. M. W. Tsui
Johns Hopkins University, Baltimore, MD, USA

*Abstract* – SPECT systems using pinhole apertures permit radiolabeled molecular distributions to be imaged in vivo in small animals. Nevertheless studying cardiovascular diseases by means of small animal models is very challenging. Specifically, submillimeter spatial resolution, good energy resolution and high sensitivity are required. We designed what we consider the "optimal" radionuclide detector system for this task. It should allow studying both detection of unstable atherosclerotic plaques and monitoring the effect of therapies. Using mice is particularly challenging in situations that require several intravenous injections of radiotracers, possibly for week or even months, in chronically ill animals. Thus, alternative routes of delivering the radiotracer in tail vein should be investigated. In this study we have performed preliminary measurements of detection of atherosclerotic plaques in genetically modified mice with high-resolution prototype detector. We have also evaluated the feasibility of assessing left ventricular perfusion by intraperitoneal delivering of MIBI-Tc in healthy mice.

## 1 Introduction

Heart failure is the leading cause of disability and mortality in the developed countries, and ischemic heart disease is the leading cause of heart failure. In most cases, ischemic heart disease is caused by atherosclerosis, usually present even when the artery lumens appear normal by angiography. The clinical challenge is not just in identifying the patient with atheroma but in recognizing specific lesions likely to cause clinical events, that means "vulnerable" plaque. For this reason the assessment of myocardial perfusion plays an important role in the diagnostic work-up of patients with heart failure as well as in the assessment of prognosis and guiding of therapy [1]. Molecular imaging by radionuclides is the most reliable non invasive technique for myocardial perfusion studies. SPECT better than PET is the technique of choice. It has been shown [2,3] that after careful calibration, using standard nuclear medicine software, perfusion ECG gated SPECT in mice permits quantification of Left Ventricular volumes and motion. This would allow evaluating the effects of therapy in the limit of the sensitivity attained by the system. In fact the magnitude of $^{99m}$Tc-MIBI uptake predicts the response of myocardium with abnormal function to subsequent revascularization in chronic coronary artery disease, and the recovery of myocardium after reperfusion therapy for acute myocardial infarction.

Studies with animals, namely mice are very important for the similarities with human coronary artery diseases. Furthermore, the wide availability of genetically and surgically induced murine models of heart failure provides an unique opportunity for relating specific molecular mechanisms to functional outcomes. However, studying cardiovascular diseases by means of small animal models is very challenging because submillimeter spatial resolution, good energy resolution and high sensitivity are required. The goal of the experiment dictates the spatial resolution and sensitivity required for the imaging system. Many devices have been proposed or developed, each of them with different performance. Most of them are based on Anger cameras with pinholes and multipinholes [4]. Due to the high efficiency and to the magnification factor that can be used good performances have been showed especially using multipinhole techniques. Nevertheless the Anger camera based systems have limitation [5] for many reasons. The ideal system should be an "open" and flexible, to be integrated in multimodality with other detectors (MRI, optical). This is not possible with the Anger camera based systems. Additionally, injecting radiotracers in mouse models of heart failure many times, possibly for week or even months, as needed, is impossible using the usual route of delivery (tail vein). Alternative routes have to be used.

This paper describes the research started by our collaboration in outlining the best detectors suited for these studies, the preliminary measurements, with an high resolution detector prototype, for the detection of atherosclerotic plaques on genetic modified mice and perfusion measurements comparing the uptake of Tc-MIBI with the two modalities: tail vein and peritoneum.

## 2 MATERIALS AND METHODS

We designed what we consider the "optimal" radionuclide detector system for this task, flexible enough to be integrated in a multimodality system: up to 8 detector heads to optimize the trade-off between spatial resolution and sensitivity. One of these dedicated modules of a detector with spatial resolution around 800 μm, sensitivity of 0.035 cps/kBq, and active area $100 \times 100$ mm$^2$, using tungsten pinhole collimator(s) and a high granularity scintillator (CsI(Na) with 0.8 mm pitch, the smallest so far achieved for SPECT detectors) or continuous LaBr$_3$.

For guidance, at submillimeter resolution, the efficiency of the single detector head is twice a typical Anger Camera for small animals.

### 2.1 Detector layout

The reasons fr the choice of the layout comes from the fact that in the SPECT with multipinholes, with 3D reconstruction, a suffcent number of "resolution elements" has to be used. This translates in the need of 120 pixels in 100 mm, that means an intrinsic spatial resolution of 0.8 mm. Scintillator arrays of very small pixels have to be used and identifying so small pixels is challenging. It will require to fully exploit the characteristics of the electronics we have designed and built, capable of reading out up to ~4096 channels individually at 20 kHz [6,7].

Preliminary test measurements with small samples (see Figure 1) show that very good pixel identification is obtained coupling such a scintillator to a PSPMT Hamamatsu H9500 ($3 \times 3$ mm$^2$ anode pixel).

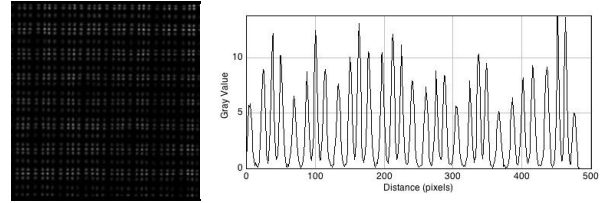

Figure 1: Flood field (Co-57) and pixel identification of CsI(Na) 0.8 mm pitch coupled to Hamamatsu H9500 (see text for details)

Nevertheless this layout could be not optimized in the dead areas between the PSPMT's [9,12]. Moreover, since good energy resolution is required in multilabeling technique, careful comparison has to be done between CsI(Tl) 0.8 mm pitch and 4 mm thin LaBr$_3$(Ce) provided that the latter can be built. This solution would allow to get rid of the dead area problem and would provide very good energy resolution.

### 2.2 Prototypes

Two prototypes have been developed and implemented: the first prototype had the best achiveable intrinsic spatial resolution at the cost of reduced FOV; it was equipped with CsI(Tl) scintillator, 1 mm pitch, $50 \times 50$ mm$^2$ size, coupled to PSPMT Hamamatsu H9500 ($3 \times 3$ mm$^2$ anode pixel). It has been used to evaluate realistic performance of an optimal detector for its intrinsic resolution (example in Figure 1).

Table 1: Components and characteristics of the second prototype.

| Pinhole diameter (mm) | 0.5 |
|---|---|
| Scintillator | NaI(Tl) |
| Pitch (mm) | 1.5 |
| Thickness (mm) | 6 |
| Area (mm2) | $100 \times 100$ |
| Photosensor | $2 \times 2$ PSPMT H8500 |
| Intrinsic Resolution (mm) | 0.8 |
| Efficiency (cps/MBq) | 35 |
| Magnification factor | 3 - 4 |
| FOV (mm2) | $33 \times 33 - 25 \times 25$ |

The second prototype had performances close to the needs in terms of spatial resolution and adequate Field Of View (FOV). The main components and characteristics of this device are reported in Table 1; it has been used for the measurements described in this paper, coupled to pinhole collimator with M=3 magnification providing a FOV of ~33×33 mm$^2$ sufficient for imaging the mouse body relevant for studying the thorax region and M=3.4 with FOV of ~25×25 mm$^2$, adequate for heart perfusion imaging. This allowed the study of the basic problems of the detection system and animal handling.

### 2.3 Micro SPECT System prototype

The prototype SPECT system consists in the detectors described above equipped with a 2.5 cm diameter acrylic cylindrical bed-holder (3 mm thick) that keeps the mouse in horizontal position (see Figure 2).

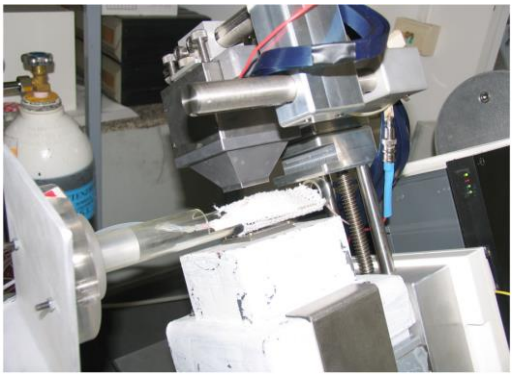

Figure 2: The SPECT system prototype: the small FOV high resolution detector above the acrylic bed-holder and larger FOV NaI(Tl) based detector head below the bed-holder.

The detectors are mounted on a motorized gantry that can rotate around a fixed bed-holder. The system could be manually adjusted to optimize the distance between the pinhole and the axis of rotation, giving the possibility to resize the camera parameters depending on measurements needs.

### 2.4 Phantom Studies

To test the tomographic spatial resolution capabilities of our detector, a miniature acrylic resolution phantom was manufactured, as shown in Figure 3. It consists of six sectors, each containing equally sized sets of 0.5 mm height capillaries with diameters and minimum distances between them from 0.8 to 1.3 mm in step of 0.1 mm. The overall phantom diameter was 2.5 cm. The single pinhole projection data were acquired in 60 angular intervals over 360 degrees at 2 min/projection. The total activity of all capillaries was ~4.5 mCi (166 MBq) of $^{99m}$Tc.

### 2.5 Animal procedures, Anaesthesia, and Tracer administration

Three adult male FVB/N mice of 12 weeks of age (Charles River Laboratories, Inc.) were used. Animal studies were crried out in compliance with the Italian guidelines for animal care (DL 116/92). The mice were intraperitoneally anesthetized with ketamine and xylazine and mechanically ventilated (respiratory rate was set at 110 breaths/min). Care was taken to minimize, as much as possible, the volume of injected radio-tracers around 0.02-0.05 ml to avoid significant changes in the whole blood volume of the small animals.

Thoracic bone scan was performed on the first mouse to evaluate system image quality (mouse was injected with 2 mCi of $^{99m}$Tc-MDP). Tomographic acquisitions started 2 hour after tracer administration. The single pinhole projection data were acquired as for the phantom: 60 angular intervals over 360 degrees, 120 s/projection.

Myocardial perfusion scan was performed on the second mouse: it was injected with 6.7 mCi of $^{99m}$Tc-MIBI; acquisitions started 1 hour after tracer administration to ensure a better contrast of heart to soft tissues. Projection data were acquired at 60 s/projection.

The same procedure was used for the third mouse but it was injected with 6.7 mCi of $^{99m}$Tc-MIBI intraperitoneally. To assure high-resolution and artefacts free SPECT image reconstruction, mechanical calibration of detector was needed. For this reason, tomographic acquisition of a set of 2-point sources (~1 mm in size) positioned as far as possible both along the axis of rotation and away of it, was also performed.

### 2.6 Image reconstruction technique

Upper head projection data were reconstructed using a 3D pinhole OS-EM algorithm [10] which takes into account system geometric misalignment parameters, including the center-of-rotation error, the tilt angles between the axis-of-rotation and the detector plane in the 3D space. Reconstruction matrix size was 90×90×90 with a voxel size of 0.25 mm. A 3D Butterworth filter was used for the post-reconstruction imaging.

### 2.7 Myocardial perfusion analysis

Since there is no true standard for quantification of SPECT [11], we used the Standardized Uptake Value, SUV, calculated as a ratio of tissue radioactivity concentration (in units kBq/ml) at time T, C(T) and injected activity (in units MBq) A at the time of injection divided by body weight (in units kg) W: $SUV = C(T)/[A/V]$.

Assuming the radiotracer uniformly distributed, and taking into account the time of measurement t relative to the injection time, we calculated the SUV as Regional Uptake Value RUV measuring the counts S in the volume of interest (heart) V: $RUV = Se^{\lambda t}/[V/C]$.

### 2.8 Detection of atherosclerotic plaques

The same detector prototype with NaI(Tl) scintillator array, has been used for a preliminary evaluation of the detection capability of atherosclerotic plaques by [99m]Tc labeled Annexin V[9] radiopharmaceutical. Male mice lacking of a functional apolipoprotein E (Apo-E) gene were used. The mice are healthy when born, but have a markedly altered plasma lipid profile compared to normal mice, and rapidly develop atherosclerotic lesions. The imaging procedure was very similar to the perfusion study: 60 angular intervals over 360 degree, 60 s/projection, acquisition started 1 hour after injection of ~3 mCi of [99m]Tc-Annexin V.

## 3 RESULTS

### 3.1 Phantom studies, Spatial Resolution, Sensitivity

For the resolution phantom as well as for myocardial perfusion studies we used a FOV of 33 mm. Figure 3 right report one view of the reconstructed resolution phantom: the 0.8 mm capillaries are cleary separated demonstrating an even better spatial resolution.

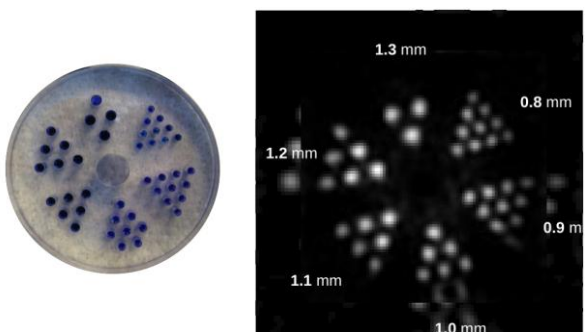

Figure 3: Miniature acrylic resolution phantom (left), and reconstructed image (right), sum of 21 trans-axial slices. 0.8 mm capillaries are clearly separated on image; peripheral artefacts can be easily masked.

The sensitivity of the system of ~35 cps/MBq was estimated using a point-like source of 370 kBq [57]Co in the centre of the FOV at a distance of 10 mm. The energy resolution of the main [57]Co photopeak was 14%.

### 3.2 Perfusion images

Figure 4 shows a perfusion image obtained with the NaI(Tl) detector prototype. Mid-ventricular short-axis slice (left) and horizontal long-axis slice (right) obtained from reconstructed projection of myocardial perfusion images are shown. Left and right ventricular cavities and corresponding walls can be easily identified. Image data were acquired with double head stationary SPECT system, during 60 minutes starting 1 hour after administration of 6.7 mCi of [99m]Tc-MIBI.

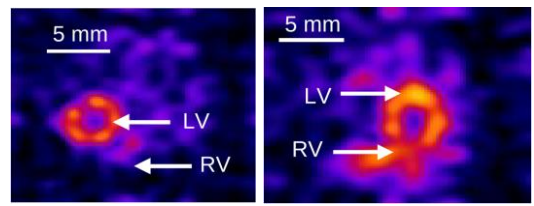

Figure 4: Short-axis (left) and horizomtal lomg axis (right) slice of a SPECT study of the mouse hearth. Short and long-axis slices show a good left ventricular wall chamber visulazation. LV, left ventricle; RV righ ventricle

The need of a different route of delivery brought us to new scan with comparison of two different modalities on two different mice, injection through the tail vein and injection through the peritoneum as mentioned above; the imaging procedure was identical. Figure 5 reports the transverse, sagittal and coronal views obtained by injecting the radiotracer into the tail vein.

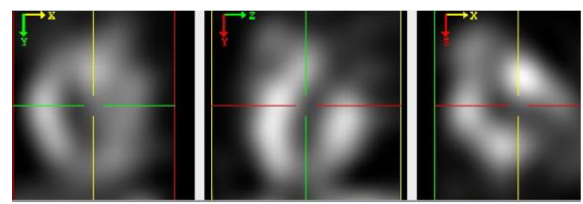

Figure 5: Transverse, sagittal and coronal heart views. Tail vein injection.

The other mouse had the radiotracer injected peritoneally: Figure 6 shows the corresponding perfusion images.

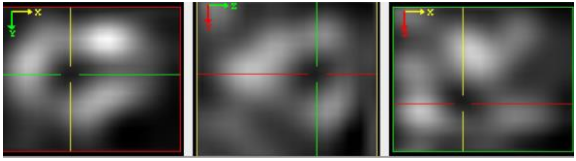

Figure 6: Same as Figure 5 for the mouse injected peritoneal.

### 3.3 Perfusion Uptake

Table 2 reports the results of the calculated uptake for the two delivery modalities.

Table 2: Quantitative data from the perfusion study by different injection modalities.

|  | Peritoneum | Tail vein |
|---|---|---|
| Activity (MBq) | 170 | 130 |
| Weight (g) | 37 | 37 |
| Age (week) | 12 | 12 |
| Time/view (m) | 1 | 1 |
| Number of Views | 60 | 60 |
| SUV |  |  |
| Transverse | 0.62 | 1.09 |
| Coronal | 0.51 | 1.22 |
| Sagittal | 0.53 | 1.31 |

A reduction of uptake occurred in peritoneum injection, but the ventricular cavities are identified in both cases.

### 3.4 Detecting atherosclerotic plaques

Mice of different ages were scanned. First results are shown in Figure 7 and Figure 8. Young mouse in Figure 7 didn't show plaques, in the limit of the sensitivity of the detector. Uptake of Annexin V[9] by liver is visible.

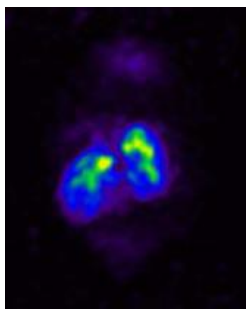

Figure 7: ApoE mouse six weeks old; visible is the liver uptake.

Figure 8 left shows the result for the control mouse. It does not present any suspicious spot. Figure 8 right shows the APOE (+/-) mouse: suspicious spots seem to be visible in the image. Hystological findings confirmed the above interpretatin of the images. No definitive conclusions can be extracted form this preliminary, and statistically limited analysis, but the adopted methodology and protocol look promising.

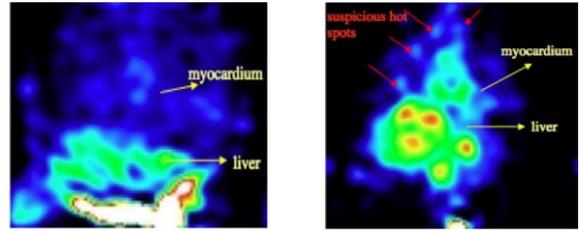

Figure 8: Twenty-five weeks old mice; left: Control mouse, right: ApoE mouse.

### 4 CONCLUSIONS AND OUTLOOK

Multi head high-resolution detector prototype has been designed as optimized SPECT system for preclinical research, specifically oriented on detection of atherosclerotic vulnerable plaques and on the evaluation of perfusion and function of the left ventricle. Two detector heads with different components have been built and used in the present study, whose scope was the estimation of the main detector performances and evaluation of the animal handling issues. The submillimetre spatial resolution of the prototype demonstrated to be adequate for perfusion studies, while the 14% energy resolution allows the use of dual tracers techniques. The relatively modest sensitivity can be increased either using more detector heads simultaneously and/or larger pin-hole diameter (up to 1.5 mm will not compromise significantly the resolution). The injection of the radiotracer through the peritoneum instead of the tail vein showed comparable good perfusion, at the price of ~1/2 reduction of the uptake by the heart; in this case the sensitivity of the system has to be increased to compensate such reduction. This can be obtained by fine-tuning the parameters: pin-hole diameter and magnification, possibly the multipinhole technique and adding as many detector head as possible to the system (up to 8 are considered in the current design). Integration of other modalities [3], essentially optical and MRI, would remarkably extend the research potential of this device. However it would require significant modifications of the layout, and of the materials and components starting with substitution of PSPMT's with Silicon Photomultipliers (SiPM) insensitive to the magnetic fields. Research in this direction is ongoing.